\documentclass[pra,twocolumn,showpacs,preprintnumbers,amsmath,amssymb]{revtex4}
\usepackage{graphicx}% Include figure files
\usepackage{dcolumn}% Align table columns on decimal point
\usepackage{bm}% bold math
\usepackage{latexsym,epsfig}

\begin{document}
\preprint{preprint}
\title{Phase separation in a two species Bose mixture}
\author{$^1$Tapan Mishra\protect\footnote[1]{E-mail: tapan@iiap.res.in},
$^2$Ramesh V. Pai\protect\footnote[2]{E-mail: rvpai@unigoa.ac.in}
and $^1$B. P. Das\protect \footnote[3]{E-mail: das@iiap.res.in}}
\affiliation{ $^1$Indian Institute of
Astrophysics, II Block, Kormangala, Bangalore, 560 034, India. \\
$^2$Department of Physics, Goa University, Taleigao Plateau, Goa 403
206, India. }
\date{\today}

\begin{abstract}
We obtain the ground state quantum phase diagram for a two species
Bose mixture in a one-dimensional optical lattice using the finite
size density matrix renormalization group(FSDMRG) method. We discuss
our results for different combinations of inter and intra species
interaction strengths with commensurate and incommensurate fillings
of the bosons. The phases we have obtained are superfluid, Mott
insulator and a novel phase separation, where the two different species
reside in spatially separate regions. The spatially separated phase
is further classified into phase separated superfluid(PS-SF) and
Mott insulator(PS-MI). The phase separation appears for all the
fillings we have considered; whenever the inter-species interaction
is slightly larger than the intra-species interactions.
\end{abstract}


\maketitle
\section{introduction}
Studies of quantum phase transitions are currently of great interest
as they provide important insights into a wide variety of many-body
systems \cite{sondhi,sachdev}. The pioneering observation of the
superfluid (SF) to Mott insulator (MI) transition in an optical
lattice using cold bosonic atoms \cite{greiner}, which had been
predicted by Jaksch et. al. \cite{jaksch}, highlights the exquisite
control of the inter atomic interactions that is possible in such
systems. In that experiment, performed using $^{87}Rb$ atoms, the
tunneling of the atoms to neighboring sites and also the strength of
the on-site interactions were controlled by tuning and/or detuning
the laser intensity in order to achieve the transition from the SF
phase (random distribution of atoms) to the MI phase where there are
a fixed number of atoms per site \cite{greiner}. Recent developments
involving the manipulation of ultracold atoms have led to the
realization of genuine one dimensional systems such as the
Tonks-Girardeau gas~\cite{paredes}. Several interesting phenomena
including the SF-MI transition have been observed in one-dimensional
optical lattices~\cite{stoferle}.

In the past few years, on the theoretical side, many investigations
have been carried out using a single species of bosonic atoms in
optical lattices~\cite{batrouni,rvpai}. Recently cold bosonic
mixtures~\cite{isacsson}, fermions~\cite{gu} and Bose-Fermi
mixtures~\cite{albus,lewenstein} in optical lattices have attracted
much attention. Mixtures of different species are very interesting
since additional phases could appear due to the inter-species
interactions~\cite{kuklov,mathey}.

In the present work, we consider a system with two species of
bosonic atoms or equivalently, bosonic atoms with two relevant
internal states. The two species shall be called $a$ and $b$ type
respectively.  The low-energy Hamiltonian is then given by the
Bose-Hubbard model for the two boson species:
\begin{eqnarray}\label{eq:ham}
\nonumber
H&=&-t^{a}\sum_{<i,j>}(a_{i}^{\dagger}a_{j}+h.c)-t^b\sum_{<i,j>}
(b_{i}^{\dagger}b_{j}+h.c) \\
\nonumber &&+\frac{U}{2}^{a}\sum_{i}
n_{i}^{a}(n_{i}^{a}-1)+\frac{U}{2}^{b}\sum_{i}n_{i}^{b}(n_{i}^{b}-1)\\
&&+U^{ab}\sum_{i}n_{i}^{a}n_{i}^{b}.
     \end{eqnarray}
Here $a_i$ ($b_i$) is bosonic annihilation operator for bosonic
atoms of $a$ ($b$) type localized on site $i$.
$n^a_i=a^\dagger_ia_i$ and $n^b_i=b^\dagger_ib_i$ are the number
operators. $t^{a}$ ($t^{b}$) and  $U^a$ ($U^b$) are the hopping
amplitudes between adjacent sites $\langle ij \rangle$ and the
on-site intra-species repulsive energies, respectively for $a$ ($b$)
type of atom. The inter-species interaction is given by $U^{ab}$.
The hopping amplitudes ($t^a$, $t^b$) and interaction parameters
($U^a$, $U^b$, $U^{ab}$) are related to depth of optical potential,
recoil energy and the scattering lengths. The ratio $U^{ab}/U^{a,b}$
can be controlled to a wide range of values~\cite{demler}
experimentally. In this work we consider inter-exchange symmetry $a
\leftrightarrow b$, implying $t^a=t^b=t$ and $U^a=U^b=U$ and study
the effect of inter-species interaction on the ground state of
model(\ref{eq:ham}) in one-dimension.  We set our energy scale by
$t=1$.

The model~(\ref{eq:ham}) has been studied earlier using the mean
field~\cite{demler}, Monte-Carlo~\cite{kuklov} and the Bosonization
methods~\cite{mathey} and this has resulted in the prediction of the
basic structure of its ground state phase diagram.  The Bosonization
study predicts phase separation (PS) for large values of the
inter-species interaction $U^{ab}$ by considering one species of
bosons to be hard core and the other to be in the intermediate to
hard core regime~\cite{mathey}. Phase separation has also been found
using a variational method based on the multi orbital best mean
field ansatz \cite{cederbaum}. However, a clear picture of the
transitions pertaining to the SF, MI and PS phases has not emerged
so far. In order to achieve this, we consider the influence of the
inter-species interaction $U^{ab}$ on these phases, by carrying out
a systematic study of its effect on the ground state of
model~(\ref{eq:ham}) in one-dimension using the finite size density
matrix renormalization method~\cite{rvpai,white}. The following part
of the paper is organized as follows. Section II contains the
details of our finite size density matrix renormalization method.
Section III  contains our results. We end with concluding remarks
in Section IV.

\section{FSDMRG calculation}
The Finite-Size Density-Matrix Renormalization Group (FSDMRG) method
has proven to be very useful in studies of one-dimensional quantum
systems\cite{rvpai,white}. The details of this method are given in a
recent review by Schollw$\ddot{o}$ck~\cite{dmrgreview}. The open
boundary condition is preferred over periodic boundary condition for
this method because the loss of accuracy which increases with the
size of the system is much less in the former than the latter. In
the conventional FSDMRG method, the lattice is first built to the
desired length ($L$) using the infinite system density-matrix
renormalization method (DMRG).  The finite size sweeping is done
only for this desired lattice size $L$. We use a slightly modified
form of the FSDMRG, where we sweep at \textit{every step} of the
procedure and not just for the case which corresponds to the largest
value of $L$. This enables one to obtain accurate correlation
functions. Furthermore, since the superfluid phase in models such as
Eq.~(\ref{eq:ham}), in $d = 1$ and at $T = 0$, is critical and has a
correlation length that diverges with the system size $L$,
finite-size effects must be eliminated by using finite-size scaling
as we show later. For this purpose, the energies and the correlation
functions, obtained from a DMRG calculation, should converge
satisfactorily for each system size $L$. It is important, therefore,
that we use the FSDMRG method as opposed to the infinite-system DMRG
method especially in the vicinities of continuous phase transitions.

In the FSDMRG method the bases used for left- and right-block Hamiltonians are
truncated by neglecting the eigenstates of the density matrix
corresponding to small eigenvalues which leads to truncation errors.
If we retain $M$ states, the density-matrix weight of the discarded
states is $P_M=\sum_{\alpha=1}^{M} (1-\omega_\alpha)$, where
$\omega_{\alpha}$ are the eigenvalues of density matrix. $P_M$
provides a convenient measure of the truncation errors. We find that
these errors depend on the order-parameter and correlation length for
a given phase. For a fixed $M$, we find very small truncation errors in the
gapped phase and the truncation errors are largest for the SF phase.
In our calculations we choose $M$ such that the truncation error is
always less than $5\times 10^{-5}$ and we find that $M = 128$ suffices.

The number of possible states per site in the model(\ref{eq:ham}) is
infinite since there can be any number of $a$ and $b$ species bosons
on a site. In a practical FSDMRG calculation we must truncate the
number of states $n_{max}$ allowed per site. The value of $n_{max}$,
of course, will depend on the on-site interaction $U$. The
smaller the value of $U$ the larger must be $n_{max}$. From our
earlier calculation\cite{rvpai} on related models, we find that
$n_{max}=4$ is sufficient for the value of $U$ considered here. This
implies, for model(\ref{eq:ham}),  $4$ states each per site for $a$
and $b$ species bosons and a total of $16$ states per site. This
corresponds to a truncation of bases of left (right) block from $16M$
to $M$ in each FSDMRG iteration.

Before proceeding further we give a brief summary of our results.
The various parameters that we calculate to study the ground state
properties of model~(\ref{eq:ham}) are the energy gap $G_L$, which
is the difference between the energies needed to add and remove one
atom from a system of atoms,i.e.,
\begin{equation}
G_L=E_L(N_a+1,N_b)+E_L(N_a-1,N_b)-2 E_L(N_a,N_b) \label{eq:gap}
\end{equation}
and the on-site density correlation function
\begin{equation}\label{eq:ni}
    \langle n^\alpha_i \rangle=\langle
\psi_{0LN_aN_b}| n^\alpha_i |\psi_{0LN_aN_b}\rangle.
\end{equation}
Here $\alpha$, is an index representing type $a$ or $b$ bosons,
$E_L(N_a,N_b)$ is the ground-state energy for a system of size $L$
with $N_a$ ($N_b$) number of $a$ ($b$) type bosons and
$|\psi_{0LN_aN_b}\rangle$ is the corresponding ground-state
wavefunction, which are obtained by the FSDMRG method. In $d = 1$,
the appearance of the MI phase is signaled by the opening up of the
gap $G_{L\to\infty}$. However, $G_{L}$ is finite for finite systems
and we must extrapolate to the $L\rightarrow \infty$ limit, which is
best done by using finite-size scaling~\cite{rvpai}. In the critical
region, i.e., SF region, the gap
\begin{equation}
  \label{eq:scaling-corr}
  G_L \approx L^{-1}f(L/\xi),
\end{equation}
where the scaling function $f(x) \sim x , \, x \to 0$ and $\xi$ is
the correlation length. $\xi \rightarrow \infty$ in the SF region.
Thus plots of $LG_L$ versus $U$, for different system sizes $L$,
consist of curves that intersect at the critical point at which the
correlation length for $L=\infty$ diverges and gap $G_\infty$
vanishes.

Defining the ratio of the inter and intra species interactions
$\Delta=U^{ab}/U$, we study the ground state of model~(\ref{eq:ham})
for  $\Delta < 1$, $\Delta=1$ and $\Delta
> 1$. The ground state exhibits some similarities as well as
differences in each of the cases. When the kinetic energy is the
dominant term in the model, the ground state is in 2SF (both $a$ and
$b$ species are in the SF phase) state for all $\Delta$. This
similarity is, however, lost when the interactions dominate. For
$\Delta \leq 1$, i.e., $U^{ab} \leq U$, the large $U$ phase is  Mott
insulator with non-zero energy gap in the ground state. This state
has an uniform local density of bosons for each species, i.e.,
$\langle n^a_i\rangle=\langle n^b_i\rangle$ for all $i$. The 2SF to
MI transition is possible only when the total density
$\rho=\rho_a+\rho_b$ is an integer. For $U^{ab} \sim U$, the 2SF-MI
transition for model~(\ref{eq:ham}) is similar to the SF-MI
transition for single species bosons with the same density of
bosons. For $\Delta > 1$ and for small values of $U$, the ground
state is a 2SF state. However, when $U$ increases, the ground state
first goes into superfluid phase with $a$ and $b$ bosons spatially
separated into different regions of the lattice. This is the case
when $\rho_a=\rho_b=1/2$. This phase may be called the phase
separated superfluid(PS-SF). There is no gap in the ground state
energy spectrum and the phase separation order parameter defined as
\begin{equation}
  \label{eq:o_ps}
  O_{PS}=\frac{1}{L}\sum_i \langle\psi_{0LN_aN_b}|
(|n^a_i - n^b_i|)|\psi_{0LN_aN_b} \rangle.
\end{equation}
is non-zero. A further increase in $U$ results in opening up of the
gap in the energy spectrum. This Mott insulator has a non-zero phase
separation order parameter and it may be called the phase separated
Mott-Insulator(PS-MI). The total local density $\langle
n_i\rangle(=\langle (n^a_i+n^b_i)\rangle)=\rho$  remain uniform
across the lattice. When the densities are different, for example
$\rho_a=1$, $\rho_b=1/2$, no PS-MI is found and the ground state has
only 2SF and PS-SF phases. When $\rho_a=1$, $\rho_b=1$ we find, for
$\Delta =1.05$, no PS-SF phase and the transition is directly from
2SF to PS-MI. We now present the details of our results.

\section{Results and Discussions}
In the absence of the inter-species interaction $U^{ab}$, model
(\ref{eq:ham}) is an independent mixture of the individual species
of bosons. So the nature of the ground state of model (\ref{eq:ham})
depends only on the density of the individual species of bosons:
$\rho_a$, $\rho_b$ and $U^a=U^b=U$, the on-site interactions. For
example, if $\rho_a\ne n$, $\rho_b\ne n$, where $n$ is an integer,
the ground state is always in the superfluid phase irrespective of
the strength of the on-site interaction $U$. The Mott insulator is
possible only when either $\rho_a=n$ or $\rho_b=n$. Based on the
values of $\rho_a$, $\rho_b$ and $U$, the ground state is
categorized as 2SF(both $a$ and $b$ type bosons in SF phase),
SF+MI($a$ boson in SF and $b$ in MI phase or vice-versa) and
MI+MI(both $a$ and $b$ bosons in MI phase). For $\rho_a\ne n$,
$\rho_b\ne n$ or for any values of $\rho_a$, $\rho_b$, but $U <
U_c$, where $U_c$ is the critical on-site interaction for SF to MI
transition, the ground state is always in the 2SF phase.  SF+MI
phase is possible for $\rho_a\ne n$, $\rho_b=n$ ( or vice-versa) and
$U > U_C$. If both $\rho_a=\rho_b=n$ and $U > U_C$, we have MI+MI
phase. In order to investigate the influence of $U^{ab}$ on these
ground states, we consider three cases: $\Delta < 1$, $\Delta=1$ and
$\Delta > 1 $, where $\Delta=U^{ab}/U$. In each of these three
cases, we consider three different ranges of densities:(i)
$\rho_a=\rho_b=1/2$, (ii) $\rho_a=1, \rho_b=1/2$, (iii)
$\rho_a=\rho_b=1$.  The choice of these three cases are made to
understand the effect of the inter-species interaction on 2SF, SF+MI
and MI+MI phases. We now discuss each case below.

\begin{figure}[htbp]
  \centering
  \epsfig{file=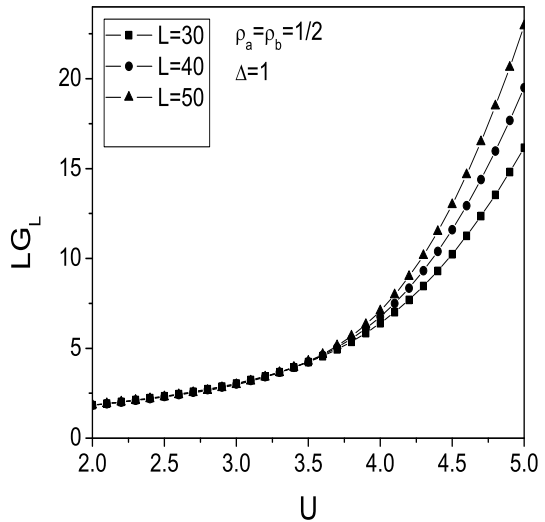,width=8cm,height=8cm}
  \caption{Scaling of gap $LG_L$ is plotted as a function of $U$ for
    different system sizes for $\rho_a=\rho_b=1/2$, $\Delta=1$.
    The coalescence of different curve for $U \le 3.4$ shows a
Kosterlitz-Thouless-type 2SF-MI transition. This transition is
similar to the SF-MI transition for single species Bose-Hubbard
model for $\rho=1$~\cite{rvpai}.} \label{fig:fig1}
\end{figure}

(i)$\rho_a=\rho_b=1/2$:\\
As discussed in the previous paragraph, for this case, there is no
MI phase if $U^{ab}=0$ and the model (\ref{eq:ham}) will have only
the 2SF phase. However, with the introduction of inter-species
interaction, the 2SF phase is destroyed. For example,
Figure~(\ref{fig:fig1}) shows a plot of scaling of gap $LG_L$ versus
$U$ for $\Delta=1$. Curves for different values of $L$ coalesce for
$U \le U_c \simeq 3.4$ indicating a gapped MI phase for $U
> U_c$. The emergence of this phase is due to the intra-species
as well as interspecies interaction strengths. The fact that
$U_c\simeq 3.4$, indicates that the model (\ref{eq:ham}) when
$\Delta = 1$ behaves like a single species of bosons at unit density
~\cite{rvpai}. These results are along the expected lines because,
when $U^{ab}= U$, every boson in the system interacts with rest of
bosons, irrespective of whether they are of type $a$ or $b$, with
the same strength and therefore the species index become irrelevant.
However, the situation changes when the inter-species interaction
$U^{ab} \ne U$.

\begin{figure}[htbp]
  \centering
  \epsfig{file=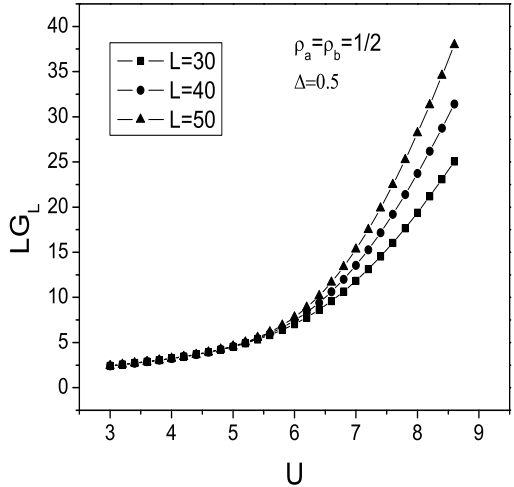,width=8cm,height=8cm}
  \caption{Scaling of gap $LG_L$ is plotted as a function of $U$ for
    different system sizes for $\rho_a=\rho_b=1/2$, $\Delta=0.5$.
    The coalescence of different curve for $U\simeq 5.4$ shows a
Kosterlitz-Thouless-type 2SF-MI transition.} \label{fig:fig2}
\end{figure}

For $\Delta < 1$, i.e, $U^{ab} < U$, the system still undergoes
2SF-MI transition when the on-site repulsion increases, but with a
higher $U_c$. For example figure~(\ref{fig:fig2}) shows a plot of
scaling of gap $LG_L$ versus $U$ for $\Delta=0.5$. The critical
$U_c(\Delta=0.5)\sim 5.4$ is substantially greater than
$U_c(\Delta=1)\sim 3.4$. The ground state of the model(\ref{eq:ham})
for $\rho_a=\rho_b=1/2$, $\Delta <1 $ consists only of 2SF and MI
phases. The transition from 2SF to MI is of
Kosterlitz-Thouless-type.

\begin{figure}[htbp]
  \centering
  \epsfig{file=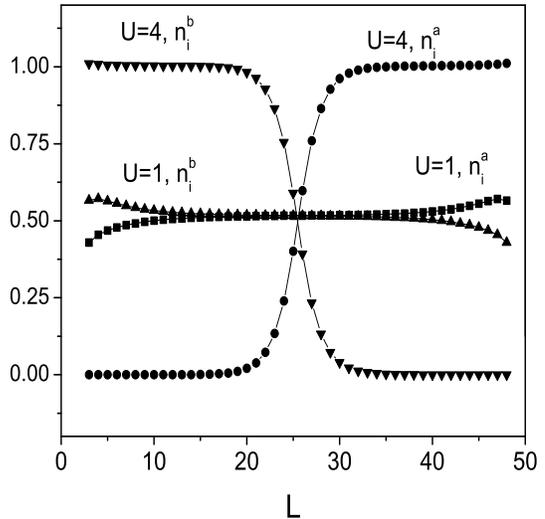,width=8cm,height=8cm}
  \caption{Plots of $\langle n^a_i\rangle$ and $\langle n^b_i\rangle$ versus $i$ for
  $U=1$ and
  $U=4$. These plots are for $\rho_a=\rho_b=1/2$, $\Delta=1.05$ and for system size $L=50$.
  The deviation in $\langle n^a_i\rangle$ and $\langle n^b_i\rangle $ near the boundaries
  for $U=1$ is due to the open
  boundary condition used
  in our FSDMRG.} \label{fig:fig3}
\end{figure}

\begin{figure}[htbp]
  \centering
  \epsfig{file=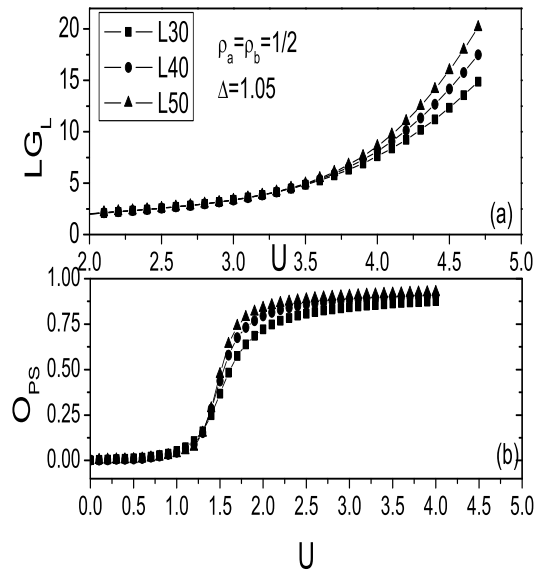,width=8cm,height=10cm}
  \caption{Plots of scaling of gap $LG_L$ (a) and order parameter for phase separation
  $O_{PS}$ (b) versus $U$ demonstrate
  various phases for the case $\rho_a=\rho_b=1/2$ and $\Delta=1.05$.
     } \label{fig:fig4}
\end{figure}

When $\Delta > 1$, the scenario is drastically different from the
one seen above. The on-site densities $\langle n^a_i\rangle$ and
$\langle n^b_i\rangle$ are plotted in Fig.~(\ref{fig:fig3}) for
$\Delta=1.05$. It is clear from this figure that there is a spatial
separation between the two different species of bosons for $U=4$ and
no spatial separation for $U=1$. This highlights a \emph{phase
separation} (PS) transition as a function of  $U$. The question then
arises whether this spatially separated phase is a superfluid or a
Mott Insulator. In order to sort this out, we plot both the scaling
of the gap $LG_L$ and the order parameter $O_{PS}$ for phase
separation in Fig.~(\ref{fig:fig4}). It is evident from these
figures that the transition to the MI phase happens at around $U_c
\simeq 3.4$ and to the spatially separated phase  around $U_c \simeq
1.3$.  The gap remains zero for $1.3 < U < 3.4$. Thus for the case
$\rho_a=\rho_b=1/2$ and $\Delta=1.05$, there are three phases: the
superfluid phase (2SF) for $U<1.3$, superfluid, but phase separated
(PS-SF) for $1.3 < U < 3.4$ and finally Mott Insulator, but again
phase separated (PS-MI) for $U > 3.4$. It should be noted that the
total density of bosons $\rho=\rho_a+\rho_b$ remain constant through
out the lattice, though bosons are space separated.  The critical
values of 2SF to PS-SF and PS-SF to PS-MI transition depends of the
on the value of $\Delta$. The detailed phase diagram in the
$\Delta-U$ plane and the nature of the different phase transitions
will be reported elsewhere.

\begin{figure}[htbp]
  \centering
  \epsfig{file=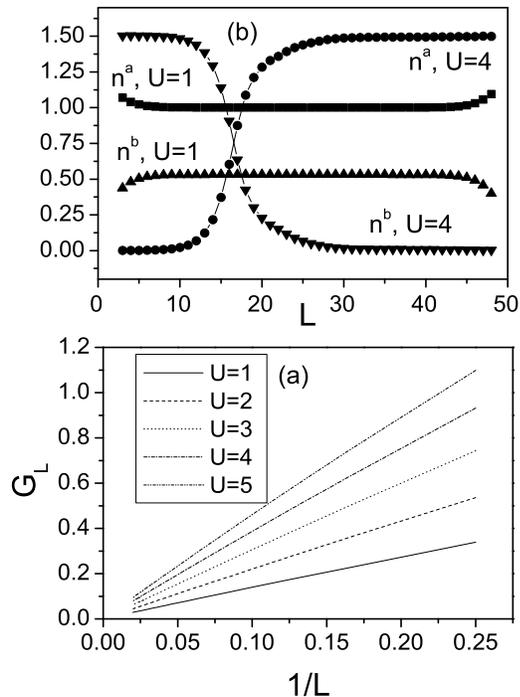,width=8cm,height=10cm}
  \caption{(a)Plots of gap $G_L$ versus $1/L$ for different values of $U$.
  The gap goes to zero linearly when $L\rightarrow \infty$
  for all the values of $U$ considered. Here $\rho_a=1$, $\rho_b=1/2$ and $\Delta=0.95$.
  (b) Local density distribution $\langle n^a_i\rangle $ and $\langle n^b_i\rangle $
  for $\rho_a=1$, $\rho_b=1/2$ and $\Delta=1.05$ for two different $U=1$, $4$.
     } \label{fig:fig5}
\end{figure}

(ii)$\rho_a=1, \rho_b=1/2$:\\
In this case, when $U^{ab}=0$  species $a$ bosons undergo a
superfluid to Mott insulator transition at $U_c \simeq 3.4$ by
virtue of having density $\rho_a=1$, while the $b$ bosons, which has
density $\rho_b=1/2$, remains in the superfluid phase. However, when
$U^{ab}\le  U$, no transition from SF to MI was found for either of
the two species of bosons. The Mott insulator phase of $a$ bosons is
completely lost. In the Fig.~(\ref{fig:fig5}(a)), we plot the length
dependence of gap $G_L$ for different $U$, which clearly indicates
that the gap vanishes at $L\rightarrow \infty$ for all values of $U$
considered. This emphasizes the fact that as far as the transition
to the Mott insulator is concerned, when $U^{ab} \le  U$, the total
density must be an integer irrespective of the densities of the
individual species of bosons; and it is this condition that really
matters.  This condition remains same for $U^{ab} > U$. Thus when
the inter-species interaction is non zero as in the present case and
the total density $\rho \ne n$, no Mott insulator phase is observed.

The phase separation, however, happens when $U^{ab} > U$. The local
density distribution of different species of bosons are given in
Fig.~(\ref{fig:fig5}(b)) for $U=1,~4$ and  $\Delta=1.05$. For $U=1$,
we find no phase separation, however, for $U=4$, the $a$ and $b$
species bosons are phase separated. They rearrange in such a manner
that the total density $\rho=\rho_a+\rho_b$ remains a constant. For
example, when $\rho_a=1$ and $\rho_b=1/2$, one-third of the region
is occupied by the species $b$ and two-third by the species $a$.  The
total density $\rho$ being $3/2$, the distributions of the $a$ and
$b$ types of bosons follow the ratio of their densities.

\begin{figure}[htbp]
  \centering
  \epsfig{file=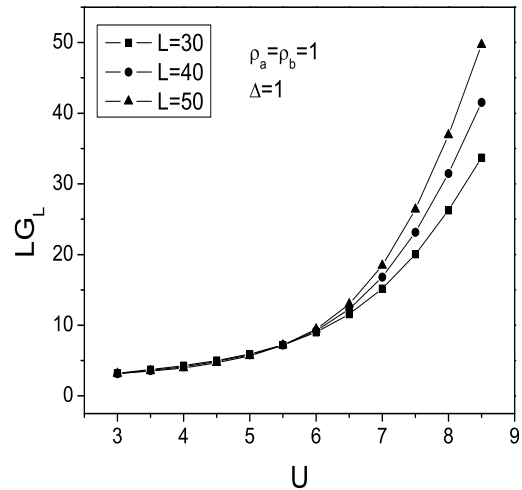,width=8cm,height=8cm}
  \caption{Scaling of gap $LG_L$ is plotted as a function of $U$ for
    different system sizes for $\rho_a=\rho_b=1$ and $\Delta=1.0$.
    The coalescence of different curve for $U\simeq 5.7$ shows a
    Kosterlitz-Thouless-type 2SF-MI transition.
     } \label{fig:fig6}
\end{figure}

(iii)$\rho_a=1, \rho_b=1$:\\
Finally we consider double commensurate case where both the species
of bosons undergo SF to MI phase transition in the absence of
$U^{ab}$. It may be noted that for $\rho_a=\rho_b=1$, $U_c\sim 3.4$
for $U^{ab}=0$.  In Fig.~(\ref{fig:fig6}), we plot the scaling of
the gap $LG_L$ for $\Delta=1.0$. From this figure and from  similar
ones for $\Delta \leq 1$ i.e., $U^{ab} \leq U$, we find that the
transition from 2SF to MI occurs at a much higher value of
$U=U_c\sim 5.7$. No SF-MI transition observed at $U \sim 3.4$. Due
to the collective intra and inter species interactions in
model~(\ref{eq:ham}), the species index become irrelevant for the
phase transition. For $U^{ab}\sim U$, the phase transition from 2SF
to MI is similar to the SF-MI transition in single species bosons
with density $\rho=2$. This is consistent with the similar
observations made  for the case of $\rho_a=\rho_b=1/2$.

\begin{figure}[htbp]
  \centering
  \epsfig{file=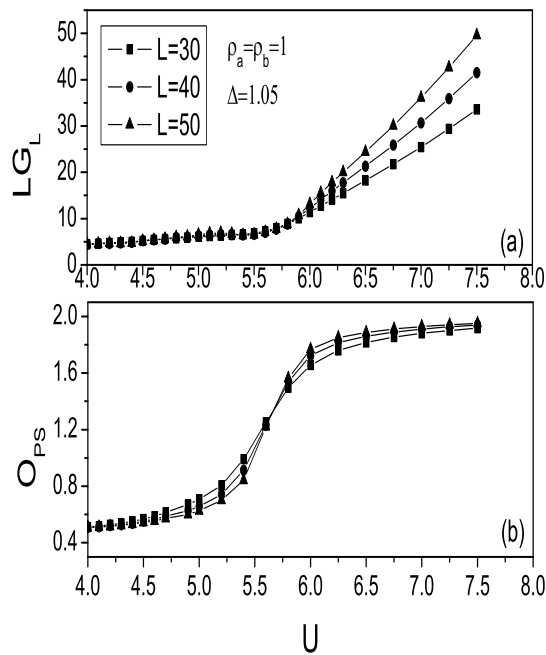,width=8cm,height=10cm}
  \caption{Plots of $LG_L$ (a) and $O_{PS}$ (b) versus $U$ demonstrate
  various phases in the case of $\rho_a=\rho_b=1$, $\Delta=1.05$.
     } \label{fig:fig7}
\end{figure}

The phase separation transition, however, occurs for $\Delta \gtrsim
1$ as given in the Fig.~(\ref{fig:fig7}). In this case the
transitions to phase separation and to the Mott insulator occur
around same $U_c\sim 5.7$. In other words we did not find a PS-SF
phase sandwiched between 2SF and PS-MI for this case.

\section{Conclusion}
We have studied the ground state of a two species Bose mixture in
one dimension using the finite-size density-matrix renormalization
group method. We have considered three sets of densities
$\{\rho_a,\rho_b\}=\{1/2,1/2\}, \{1,1/2\}, \{1,1\}$. Analyzing the
scaling of gap in the energy spectrum and the order parameter for
phase separation we have obtained several phases: 2SF, MI, PS-SF and
PS-MI. For $U^{ab} \le U$, the Mott Insulator phase is possible only
when the total density $\rho=\rho_a+\rho_b=n$, is an integer. The
superfluid to Mott Insulator transition in model~(\ref{eq:ham}) is
then similar to the single species Bose-Hubbard model with the same
total density $\rho$. The critical on-site interaction for the
2SF-MI transition, however, depends on the values of $\Delta$. The
lower the value of $\Delta$, the larger the value of $U_c$. For
$\rho \ne n$, Mott insulator phase is not found. Phase separation
occurs for  $U^{ab} > U$ irrespective of the value of density. For
$\rho=n$ and for all the values of $\Delta$ that we have considered,
we found phase separated Mott insulator phase. In the case of
$\rho_a=\rho_b=1/2$, we observe a phase separated superfluid PS-SF
sandwiched between 2SF and PS-MI. However, for $\rho_a=\rho_b=1$, no
PS-SF was found and the transition is directly from 2SF to PS-MI.
For $\rho_a=1,\rho_b=1/2$ we found a transition from 2SF to PS-SF
for $\Delta > 1$ and only 2SF phase for $\Delta \leq 1$. It would
indeed be worthwhile to devise experiments to test our findings.

\section{acknowledgments} One of us (RVP) thanks the Indian Institute
of Astrophysics, Bangalore for hospitality during the time when a
part of this work was done, DST-FIST for financial assistance and R.
Pandit and K. Sheshadri for useful discussions.
%\end{acknowledgments}

\end {document}